\documentclass{aastex}
\usepackage{emulateapj5}
\usepackage{onecolfloatx}

\def\et{{\it et al.}}

\shorttitle{New Limits on CMB Polarization}
\shortauthors{Hedman \et}

\begin{document}

\twocolumn[  %comment out if not using emulateapj

\title{New Limits on the Polarized Anisotropy of the \\ 
Cosmic Microwave Background at Subdegree Angular Scales} 

\author{M.~M.~Hedman\altaffilmark{1,4}, D.~Barkats\altaffilmark{1},
J~O.~Gundersen\altaffilmark{2},
J.~J.~McMahon\altaffilmark{1},\\ S.~T.~Staggs\altaffilmark{1},
B.~Winstein\altaffilmark{3}} 

\affil{$^1$ Department of Physics, Princeton University, Princeton, NJ, 08544}
\affil{$^2$ Department of Physics, University of Miami, Miami, FL, 33146}
\affil{$^3$ Center for Cosmological Physics and Enrico Fermi Institute
and Department of Physics, University of Chicago, Chicago, IL, 60637}

\begin{abstract}

We update the limit from the 90~GHz PIQUE ground-based polarimeter on the
magnitude of any polarized anisotropy of the cosmic microwave
radiation. With a second year of data, we have now limited both Q and
U on a ring of $1^\circ$ radius. The window functions are broad: for
E-mode polarization, the effective $\ell$ is $\left<\ell_E\right> =
191_{-132}^{+143}$. We find that the E-mode signal can be no greater
than $8.4~\mu$K (95\% CL), assuming no B-mode polarization.  Limits on
a possible B-mode signal are also presented.

\end{abstract}

]  %comment out if not using emulateapj

\altaffiltext{4}{Present address: Center for Cosmological Physics and
Enrico Fermi Institute, University of Chicago, Chicago, IL, 60637} 

\section{INTRODUCTION}
\nopagebreak

The current limits on cosmic microwave background (CMB) polarization
restrict the amplitude of its fluctuations to less than $10~\mu$K
at 95\% CL. At large angular scales, Keating \et~(2001) limit the amplitude to 
$8~\mu$K. At subdegree angular 
scales, the constraint from Hedman \et~(2001, hereafter H01)  is
$10~\mu$K, while at 
arcminute scales, Subrahmanyan \et~(2000) 
set a limit of $10~\mu$K. Present estimates of the peak
polarized fluctuation amplitude are $\sim 6~\mu$K at an angular scale
of $\sim 0\fdg2$ ($\ell\!\sim\!950$).
These estimates are based on parameters gleaned from CMB temperature anisotropy
measurements (e.g. Pryke \et~2002; Jaffe \et~2001; Wang, Tegmark, \&
Zaldarriaga 2001). While CMB
polarization has yet to be detected, its characterization will
complement CMB temperature anisotropy data and impact our
understanding of:
gravitational waves from the inflationary epoch (e.g. Turner
1997, Caldwell, Kamionkowski, \& Wadley 1999); peculiar velocities at
the surface of last scattering \citep{zalhar95}; the nature of primordial
perturbations (e.g. Spergel \& Zaldarriaga 1997); primordial magnetic
fields \citep{kosloe96}; and cosmological parity violation 
(Lue, Wang, \& Kamionkowski 1999). Here, we report improved limits
derived from new data from the 2001 observing season of the Princeton IQU
Experiment 
%\footnote{PIQUE  measured three  Stokes parameters: I, Q, and U.} 
(PIQUE) at  90~GHz.  We combine the
new data with data from the first observing season, and also present a
reanalysis of those earlier data. These data pass extensive checks for
systematic contamination.  Future publications will report 
results from a 40~GHz polarimeter also  deployed during the 2001
observing season, and give details of the instrument.

\section{INSTRUMENT, OBSERVATIONS AND CALIBRATION}
\nopagebreak

PIQUE has been described previously (H01). The results reported here
are from PIQUE's broadband 90~GHz
correlation polarimeter, which 
underilluminates a 1.2~m off-axis parabola \citep{wol97}, resulting in 
a beamsize of $0\fdg 235$. The  84-100~GHz bandpass is divided into three
subbands called S0, S1, and S2 (H01). Observations are made of 
a ring of radius $1^\circ$ around the NCP; the telescope site is 
Princeton, NJ.

The polarimeter observed the sky from 
2000 January 19 to 2000 April 2 and from 2000 December 19 
to 2001 February 28. These two observing seasons yielded 810 hrs and 
660 hrs of raw data, respectively.

The scanning strategy is designed to permit null tests for checking 
sensitivity to systematics.  
During both observing seasons, the telescope alternated between 
two azimuth positions at fixed elevation.  Data from the two 
positions are differenced to remove sensitivity to DC offsets.
For the first season 
these azimuth positions were $\pm 0\fdg 93$, and the elevation 
was $41\fdg 0$. The telescope therefore
measured $\mp Q$ (as defined by the IAU) for two regions 
separated by six hours in right ascension (RA) on the 
ring of declination $89^\circ$.  For the second season the azimuth
positions were $\pm 1\fdg 31$, with elevation $40\fdg 3$, so
the telescope measured $+U$ for two regions 
separated by twelve hours in RA on the same 
ring.  The azimuth chop period was 13 seconds until 2001 January 23, 
at which point it was doubled.

The polarimetry channels are calibrated to 10\% using a nutating aluminum flat
(H01, Staggs \et~2002). Constant elevation scans of Jupiter
are used to   
determine pointing accuracy and map the beams. For the second observing 
season the measured beam FWHM are $0\fdg 235(7)$ in co-elevation and
$0\fdg 233(7)$
in elevation, in agreement with measurements from the first season.
The absolute pointing offsets in elevation are smaller than
$0\fdg 03$.  However, early in the second observing season, the 
encoder suffered a misalignment so that the azimuth offset increased 
from $+0\fdg 03$ to $+0\fdg 08$. (No concomitant change in
beamshape was observed.) Note that we are able to neglect this in the
analysis because the overlap between the ideal and misaligned beams is
still 85\%. In fact, simulations
indicate the misalignment has less than a $2\%$ effect on our derived limit. 

\begin{figure}[ht]
%\epsscale{0.9}
\plotone{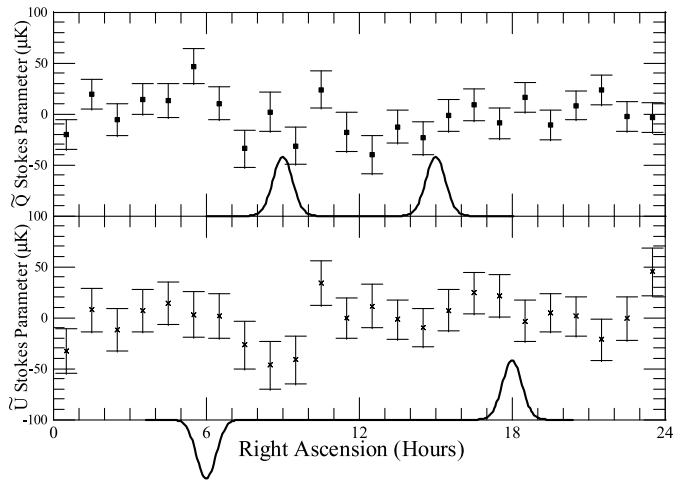}         
\caption{\small 
Binned data are shown in thermodynamic units for each observation
season.  The three frequency channels have been coadded and the one
sigma errors include small correlations among the channels.  For
clarity only 24 approximately beam-sized bins are shown: the
actual analysis uses 144 independent bins.
Due to PIQUE's differencing strategies, a  1~Jy point source with
20\% polarization located at  
$\alpha=12^h$, $\delta=89^{\circ}$ generates
the point source responses shown in the
two panels.  No such point source is known, and  
Toffolatti \et~(1998) estimate there are fewer than 200 point sources 
brighter than 1~Jy in the entire sky at 90~GHz.}  
\label{fig:datalike}
\end{figure}

\section {DATA REDUCTION} 
\nopagebreak

The 1470 hours of data from the two observing seasons include: 383 hours
of data taken while the telescope was slewing between the desired scan 
positions, 339 hours of data corrupted by known electromechanical failures, 
and 59 hours of data in isolated fragments less than 12 hours in length. 
The remaining 685 hours of data contain 186 hours of data corrupted by 
meteorological phenomena (clouds), which are identified using the 
selection criteria described below.

As discussed in H01, the correlation polarimeter suffers a small
sensitivity ($\la 0.5$\%) to total power because signals reflected
from the input of the amplifier in one arm can couple into the other
arm through the orthomode transducer.  Therefore, the distributions of
correlation coefficients
\begin{equation}
{\cal C}_N=\frac{<S_i S_j>}{<S_i^2><S_j^2>}
\end{equation}
between pairs of 
polarimetry channels $(i,j)$ display large
positive tails due to periods of rapid atmospheric fluctuations.
Data corrupted by clouds are removed by requiring the coefficients to
be less than certain thresholds.  Such selection criteria are
determined based upon a data set designed to be insensitive to real
astronomical 
polarized signals: the quadrature data (for
which data from each scan position are split into halves and 
differenced). 
The ${\cal C}_N$ are generated  as averages over $N$ chops; varying $N$
varies the time scale probed.  For purely Gaussian noise, the shapes of the 
resulting distributions of ${\cal C}_N$ for the whole data set depend
only on $N$. In order to avoid using additional
cutting measures (such as the 6-hr null test used in H01), the 
${\cal C}_N$-selection technique has been refined from that
used for H01.  Here, two time scales are used rather than one.
First we calculate
the average correlation coefficients for segments of the time series 
40-70 minutes long (specifically, $N=200$).  Segments with 
coefficients larger than a threshold of 0.20 are removed. 
Next, coefficients are 
calculated for $N=50$, for which the cut threshold is 
0.32.  These thresholds are at $2.5\sigma$ and $2\sigma$, 
and are chosen so that either cut alone removes $\sim 20$\% of the
data. The combined cuts remove  $ 27$\% of the data. Null test  
results are not sensitive to the exact values for the thresholds.
  
The 307(192) hours of data surviving the cuts detailed above
from the first(second)
observing season are parsed into 144 bins based on the Local Sidereal Time
(LST) when the data were taken, following the same procedures outlined in 
H01. These data are plotted in Figure~\ref{fig:datalike}. Offsets on
the order of a few hundred $\mu$K are	
removed from each
polarimetry channel for each ``deployment" (a  period of $> 12$ hours 
bracketed by periods when the instrument was tarped). The results are
not sensitive to the exact number of offsets removed.
Table~\ref{ta:chisq} presents
the results of the null tests described in H01, using the new 
selection criteria.  The $\chi^2$ distribution of these null tests is 
consistent with noise, demonstrating that
the data do not suffer from residual atmospheric contamination.

\begin{deluxetable}{llrrr}
\tablecaption{Results of $\chi^2$ consistency tests.\label{ta:chisq}} 
\tablecolumns{5}
\tabletypesize{\small}
\tablewidth{3truein} 
\tablehead{ \colhead{Year}
& \colhead{Test} & \colhead{\phantom{0000}S0\tablenotemark{a}} 
& \colhead{\phantom{0000}S1\tablenotemark{a}} 
& \colhead{\phantom{0000}S2\tablenotemark{a}} \\
} 
\startdata 
&Quad\tablenotemark{b}  & 0.96 & 0.25 & 0.19 \\ 
 &H1-H2\tablenotemark{c} & 0.18  & 0.14 & 0.33 \\ 
\raisebox{1.5ex}[0pt]{2000} &Pattern\tablenotemark{d} & 0.46 & 0.26 & 0.14 \\ 
% & \colhead{S0-S1} &\colhead{S0-S2} & \colhead{S1-S2} \\ 
%$S_i-S_j$
&Si-Sj\tablenotemark{e} & 0.62 & 0.83 & 0.83 \\  \hline

&Quad\tablenotemark{b}  & 0.38 & 0.61 & 0.28  \\ 
&H1-H2\tablenotemark{c} & 0.17 & 0.61 & 0.58 \\ 
\raisebox{1.5ex}[0pt]{2001}&Pattern\tablenotemark{d} & 0.80 & 0.12 & 0.38  \\ 
% & \colhead{S0-S1} &\colhead{S0-S2} & \colhead{S1-S2} \\ 
%$S_i-S_j$ 
&Si-Sj\tablenotemark{e}& 0.47 & 0.66 & 0.04\\ 
\enddata 
\tablenotetext{a}{\footnotesize Each numerical entry gives the 
probability of exceeding the $\chi^2$ for the  given frequency 
channel.}
\tablenotetext{b}{\footnotesize
 The quadrature test uses data from each scan position (east and west) split into two halves and differenced to yield the quantity
$(E_1-E_2)/2+(W_1-W_2)/2$.} 
\tablenotetext{c}{\footnotesize Data from the second half of each 
season are subtracted from the first half.}
\tablenotetext{d}{\footnotesize Pattern nulls are generalizations of
the 6-hour null test 
from H01, and are data sets constructed from the various
combinations of the data that should be zero given the differencing
scan strategy. If $d_t$ is the measured signal at Local Sidereal Time
$t$ in hours,
then for 2000 these combinations are $d_t-d_{t+6}+d_{t+12}-d_{t+18}$,
while for 2001 these combinations are $d_t+d_{t+12}$.}
\tablenotetext{e}{\footnotesize  Data differenced between two 
channels. The column entries are S0-S1, S0-S2 and S1-S2.}

\end{deluxetable} 

\begin{figure}
\plotone{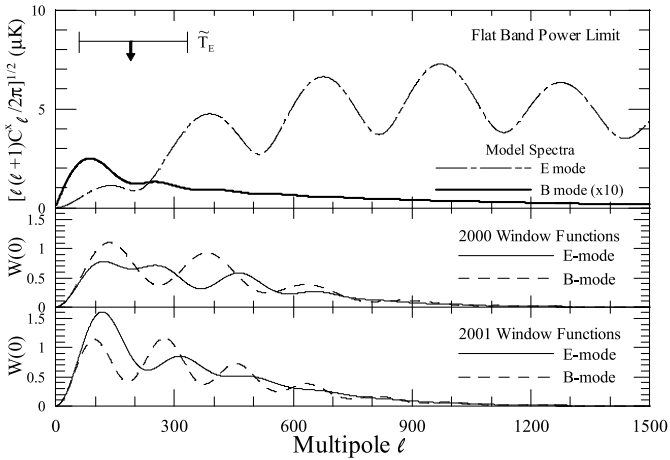}
\caption{\small
Zero-lag window functions for E- and B-modes (solid and dashed lines)
are shown in the middle panel for the 2000 observation season and in
the lower panel for the 2001 season.  The top panel shows
the limit  on $T_E$ assuming $T_B\equiv 0$ for the combined W-band
observations.  
For comparison we also plot (dashed line) E-mode predictions from the best-fit
model in Pryke \et~(2002), as well as B-mode predictions (shown in the
top panel multiplied by ten) assuming the same model with $T/S=1$.}
\label{fig:jogmega}
\end{figure}

\section{DATA ANALYSIS}
\nopagebreak

The likelihood of a model given a data vector ${\bf x}$ is 
${\cal
L}\propto\exp(-{\bf{x}}^T{\bf{C}}^{-1}{\bf{x}}/2)/|{\bf{C}}|^{1/2}$,
where  
the covariance matrix ${\bf C}$ sums both theoretical correlations
from the model (signal)  and correlations specific to the experiment 
(noise): ${\bf C} = {\bf S} + {\bf N}$. 
The analysis of the combined 2000-2001 data uses ${\bf x}$ with 
864 elements 
comprising 144 spatial pixels measured in three different 
frequency bands over the course of two observing seasons. 
The noise matrix ${\bf N}$ encodes the variances for each 
pixel for each channel and also accounts for interchannel 
correlations from both atmospheric fluctuations weakly coupled into the 
polarimeter channels and correlated gain fluctuations in the 
cryogenic amplifiers.
The interchannel correlation coefficients are $\le 8$\%
on average, and smaller for the two most sensitive channels. (S2
has just  $20$\%  of the total weight.) 
The noise matrix does not include pixel-pixel correlations, since
no such correlations are observed in the time series data.
An independent analysis using a 288-element data vector, for which
data from the three 
frequency channels are combined for each spatial pixel, with errors 
calculated to account for the interchannel correlations, yields
consistent results. 
% the likelihood figure was here and STS moved it to later.
%this table used to be later in the text.  STS.
\begin{deluxetable}{clrrr}
\tablecaption{$95\%$ Confidence Limits \label{ta:likes}}
\tablecolumns{5}
\tabletypesize{\footnotesize}
\tablewidth{3truein} 
\tablehead{\colhead{Year}&\colhead{{Data}\tablenotemark{a}} &
\colhead{$\widetilde{T}_E$\tablenotemark{b}}&
\colhead{\phm{00}$\widehat{T}_E$\tablenotemark{c}} &
\colhead{\phm{00}$\widehat{T}_B$\tablenotemark{c}} \\ 
\colhead{} & \colhead{} &\colhead{$(\mu$K)} & \colhead{\phm{00}$(\mu$K)} &
\colhead{\phm{00}$(\mu$K)} \\
} 

\startdata 
&CMB  & 12.7 & 15.8 &14.7 \\ 
&Quad & 10.3  & 13.9 & 12.9 \\ 
\raisebox{1.5ex}[0pt]{2000} 
&(H1-H2)/2 & 14.9 &19.7 &17.4 \\  \hline

&CMB  & 10.4 &15.9 &17.8 \\ 
 &Quad & 10.4  & 16.2 &18.0 \\ 
\raisebox{1.5ex}[0pt]{2001}
&(H1-H2)/2 & 11.2 &17.2 &19.4 \\  \hline
 
2000 &CMB  & 8.4 & 11.2 &11.5 \\ 
$+$ &Quad & 6.1  & 8.3 & 8.4 \\  
2001&(H1-H2)/2 & 8.5 &11.5 & 11.6 \\  \hline
\enddata

\tablenotetext{a}{\footnotesize The two null tests are
described in Table~\ref{ta:chisq}.}
\tablenotetext{b}{\footnotesize The limit $\widetilde{T}_E$ is found 
assuming $T_B\equiv0$.}
\tablenotetext{c}{\footnotesize The limits $\widehat{T}_E$ and 
$\widehat{T}_B$ are determined simultaneously 
by finding the contour of constant likelihood enclosing
95\% of the volume. }

\end{deluxetable}

Since PIQUE measured $Q$ in 2000 and $U$ in 2001,  
the signal covariance matrix takes the  form:
\begin{equation}
{\bf{S}}= \pmatrix{\langle\tilde{Q}\tilde{Q}\rangle & 
\langle\tilde{Q}\tilde{U}\rangle \cr 
\langle\tilde{Q}\tilde{U}\rangle & 
\langle\tilde{U}\tilde{U}\rangle}
\end{equation}
where 
$\langle\tilde{Q}_i\tilde{Q}_j\rangle$
(or $\langle\tilde{U}_i\tilde{U}_j\rangle$) represents the 
theoretical correlation between two spatial pixels from the 2000 
(or 2001) data 
set, and $\langle\tilde{Q}_i\tilde{U}_j\rangle$ encodes correlations 
between pixels from different years. 
Note that the $\tilde{Q}$ are sums of Q separated by six hours in  
RA and the  $\tilde{U}$ are differences of U  separated 
by 12 hours in RA.
The expression for 
$ \langle\tilde{Q}_i\tilde{Q}_j\rangle$ is given in 
H01.  Following Zaldarriaga (1998), the expression for 
$\langle\tilde{Q}_i\tilde{U}_j\rangle$ is given in terms 
of the E- and B-mode angular power spectra $C^E_\ell$ and $C^B_\ell$ by

\begin{equation}
 \langle\tilde{Q}_i\tilde{U}_j\rangle  =
 \sum_{\ell m}{ \frac{(2\ell +1)}{4\pi}}[C_\ell^E 
+C_\ell^B]W_{12,\ell}(\phi_{ij}),
\end{equation}
where
$\phi_{ij}=\cos^{-1}(\hat{n}_i\cdot \hat{n}_j)$ is the lag.
The window function $W_{12,\ell}$ has the form:
\begin{equation}
    W_{12,\ell}=\sum_m (B_{lm}^{QU})^2F_{1,\ell m}
F_{2,\ell m}\cos(m\phi_{ij}),
\end{equation}
where the $F_{\{1,2\},lm}$ are given  in terms
of associated Legendre polynomials evaluated at the ring radius 
$\theta=1^{\circ}$.
Here, the beam function is:
\begin{equation}
(B_{\ell
m}^{QU})^2=4\sin\left(\frac{m\pi}{2}\right)\cos\left(\frac{m\pi}{4}\right)\frac{\sin^2(m\pi/144)}{(m\pi/144)^2} 
e^{-\ell(\ell+1)\sigma^2},
\end{equation}
where $\sigma=0\fdg 10$ for the PIQUE beams.
Similarly, the expression for $\langle\tilde{U}_i\tilde{U}_j\rangle$
is:
\begin{equation} 
 \langle\tilde{U}_i\tilde{U}_j\rangle =
	 	\sum_{\ell}\frac{(2\ell +1)}{4\pi}[
			C_\ell^E W_{2,\ell} (\phi_{ij}) +
			C_\ell^B W_{1,\ell}(\phi_{ij})],
\end{equation}
where the  $W_{1,\ell}$ and $W_{2,\ell}$ are the associated window
functions  given in H01. In this case, the beam function is given by
\begin{equation}
(B_{lm}^{UU})^2=(B_{lm}^{QU})^2\frac{\sin(m\pi/2)}{\cos(m\pi/4)}.
\end{equation}
This formalism allows for
combination of the two data sets; however, separate analyses of 
the $\tilde{Q}$ data and the $\tilde{U}$ data are
also presented. We plot the zero-lag window functions in
Figure~\ref{fig:jogmega}. 

%The likelihood figure used to be earlier in the text. STS.
\begin{figure}
\plotone{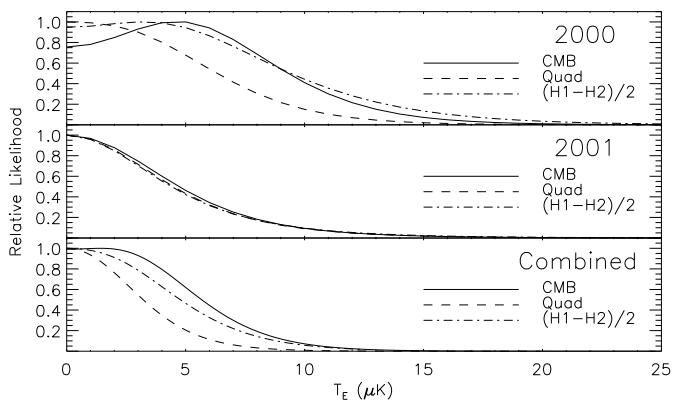}
\caption{\small
Normalized likelihoods versus flat-bandpower $T_E$ under the 
assumption $T_B=0$, as described in 
the text, for each year of data, and for the combined data.
In each panel, likelihoods for the CMB data are plotted 
along with likelihoods for two of the null data sets described in 
Table~\ref{ta:chisq}. 
}
\label{fig:likes}
\end{figure}

\setcounter{footnote}{0}

The likelihood analysis proceeds by considering flat
angular spectra, such that $\ell(\ell+1)C_\ell^X/2\pi = T_X^2$,
where $X=E, B$. Since the
amplitude of $C^B_\ell$ is predicted to be much smaller than that
of $C^E_\ell$, we first find the limit $\widetilde{T}_E$ on 
$T_E$  under the assumption $T_B$ is identically zero\footnote{The 
limit $\widetilde{T}_E$ is found by
integrating ${\cal L}(T_E,0)$; the result is $\approx 30$\% higher
if ${\cal L}(T_E^2,0)$ is integrated.}. This limit is compared to
predictions in Figure~\ref{fig:jogmega}. Next, joint
upper limits $(\widehat{T}_E, \widehat{T}_B)$ are determined by finding the
constant contour of ${\cal L}(T_E,T_B)$ enclosing 95\% of the
volume of ${\cal L}$.  This is repeated for each year separately and for 
the
combination of the two years.  The 95\% confidence level (CL) upper limits 
are
shown in Table~\ref{ta:likes} and the normalized likelihoods are shown in
Figure~\ref{fig:likes}.
The null data sets are treated in an identical manner; results are 
tabulated in Table~\ref{ta:likes} and plotted in 
Figure~\ref{fig:likes}. The limits in Table~\ref{ta:likes} do not include 
calibration
errors.  Note that for PIQUE's broad window functions, 
the 5\% beam errors only add 2\% errors in quadrature with the 10\%
calibration errors.

%The confidence limits table used to be here.  STS.

\section{DISCUSSION}
\nopagebreak

The main result here is a new constraint on the amount of
polarized anisotropy in the CMB at sub-degree angular scales. 
The result derives from combining data on Q from our first
campaign with U from our second; in so doing, important
information  on the Q-U cross-correlation is included.  
We have summarized this result as a 95\% CL limit of $8.4~\mu$K on
E-modes. Given PIQUE's window functions and current theoretical
predictions (Figure~\ref{fig:likes}),  we might expect a signal of a few
$\mu$K.  The likelihood for our CMB data (Figure 3, bottom panel)
is consistent with this expectation.  When we fit to an
offset lognormal distribution (Bond, Jaffe, \& Knox 2000), we find a central
value of $2.2~\mu\mbox{K}^2$, a variance of
$(16~\mu\mbox{K}^2)^2$ and a noise-related offset of 
$18~\mu\mbox{K}^2$. 

The results presented here have been checked with two independent
likelihood analyses and supported by extensive simulations.  Our new
selection criteria, as described above: are better able to deal with
instrumental effects on different time scales;  allow us to discard
the 6-hr null test criterion we previously used; and work for both
data sets together.  However, applying the new criteria just 
to year 2000 data, we find a weaker 95\% CL limit on polarized CMB 
anisotropy than in H01: 12.6~$\mu$K rather than 10.3~$\mu$K.

We have performed a variety of simulations to address the probability of
such a change.  Recall that our result for the year 2000 data was 
essentially unchanged by relaxing the null test cut and allowing in 80 
extra hours of data, for a total of 330 hours.
For these simulations, we start with roughly a
330 hour data set which is pure noise, generated assuming the actual
weights for each period of the data set.  We then investigate how cutting
the data can change the derived limit.  From this we find
that although a) the probability of the limit not worsening when 330 hours is
reduced to 250 hours is only about 1\%,
b) a $10 ~\mu$K error from 330 hours of data is within one standard
deviation of what is expected from pure noise  
(given our experimental weights); and
c) the expected change in the limit from removing chunks of the
data to get to 300 hours of data, (our final sample for year 2000), is about
$+1.1~\mu$K with a standard deviation of about $1.6~\mu$K.  Thus our
observed change is again within one standard deviation of what is
expected.

We thus conclude that fluctuations alone can account for the change in 
the limit derived from the year 2000 data under different selection 
criteria.  This is
so even though the former cut on the null test, which we have shown is not
needed, was probably too restrictive.

The expected signal, even including foregrounds, is still smaller than our
new limit.  Given the multipole range probed by PIQUE, this result 
provides the tightest constraint yet on the polariation spectrum 
predicted from primordial density fluctuations.  

\acknowledgements

We thank Norman Jarosik, Lyman Page,
and David Wilkinson for helpful discussions and
Al Dietrich for mechanical contributions. We are grateful to Sami Amasha
and Liam Fitzpatrick for assistance with programming.
We also thank Marian Pospieszalski and the NRAO for supplying the
HEMT amplifers. Data (including the correlation matrix and
likelihood functions) will be made public upon publication of
this Letter.

This work was supported by a NIST precision measurement grant \#NANB8D0061,
by NSF grants \#PHY96-00015, \#PHY99-84440, \#PHY-0114422, and by a Sloan 
Fellowship (to STS). 

%\bibliographystyle{apj1c}
%\bibliography{paper2-b}

\end{document}